\documentclass[sigconf]{acmart}
\AtBeginDocument{%
  }

\copyrightyear{2026}
\acmYear{2026}
\setcopyright{cc}
\setcctype{by}
\acmConference[SIGSPATIAL '26]{The 34th ACM International Conference on Advances in Geographic Information Systems}{November 03--06, 2026}{Riverside, CA, USA}
\acmBooktitle{The 34th ACM International Conference on Advances in Geographic Information Systems (SIGSPATIAL '26), November 03--06, 2026, Riverside, CA, USA}
\acmDOI{10.1145/3841645.3843101}
\acmISBN{979-8-4007-2950-8/2026/11}

\begin{document}

%%
%% end of the preamble, start of the body of the document source.
\title[Accelerating Point-in-Polygon Predicates via Algebraic Hash-Joins and Discrete Global Grids at Scale]{Accelerating Point-in-Polygon Predicates via Algebraic Hash-Joins and Discrete Global Grids at Scale}
\subtitle{An Interactive Benchmark}

\author{Levente Juhász}
\correspondingauthor
\email{levente.juhasz@ufl.edu}
\orcid{1234-5678-9012}
\affiliation{%
  \institution{University of Florida}
  \city{Davie}
  \state{Florida}
  \country{USA}
}

\renewcommand{\shortauthors}{L. Juhász}

\begin{abstract}
Traditional vector-based point-in-polygon queries rely on computationally expensive geometric predicates that scale poorly for massive datasets, even when accelerated by spatial indices. Discrete Global Grid Systems (DGGS) offer a scalable alternative by discretizing geometries into hierarchical cells, transforming complex spatial relations into constant-time relational hash-joins. However, adopting a DGGS introduces an overhead to encode data, and current grid implementations exhibit a significant performance ``tooling gap.'' In this demonstration, we present an interactive dashboard that empirically evaluates these computational tradeoffs across four DGGS implementations (H3, S2, A5, and ISEA4H) using DuckDB. Through progressive scenarios, the platform visualizes the overhead of on-the-fly encoding and demonstrates how pre-indexing spatial datasets eliminates this overhead. Ultimately, the demo proves that when data is pre-indexed, all DGGS regardless of their mathematical complexity or tooling converge to sub-second join latencies, unlocking the throughput of modern vectorized execution engines.
\end{abstract}




%\received{28 June 2026}
%\received[revised]{XX}
%\received[accepted]{XX}

\maketitle

\section{Introduction}

The exponential growth of available spatial data has exposed computational bottlenecks in traditional vector-based geoprocessing architectures. In spatial relational databases, operations such as point-in-polygon queries rely on continuous coordinate geometry. To manage large datasets, these systems typically utilize spatial indices (e.g., R-Trees \cite{guttman_r-trees_1984}) to cull non-intersecting geometries via bounding-box overlap. However, spatial indices only serve as a preliminary filter and they do not accelerate the core geometric predicate evaluation. For every point that falls within a candidate polygon's bounding box, the system must execute computationally expensive ray-casting \cite{shimrat_algorithm_1962} or winding-number algorithms \cite{alciatore_winding_1995} to determine true containment. Against complex vector boundaries at scale, these geometric predicates introduce significant latency.

Discrete Global Grid Systems (DGGS) offer an alternative spatial data model by discretizing the continuous surface of the Earth into cells \cite{sahr_geodesic_2003}. By encoding continuous coordinate geometries (latitude and longitude) into discrete identifiers, DGGS transforms the geometric problem of spatial intersection into an algebraic hash-join. This shift replaces variable-time geometric calculations with constant-time $O(1)$ database operations, which have been shown to drastically improve computational scalability for large-scale geospatial operations \cite{law_using_2025}. However, adopting DGGS introduces a computational tradeoff: the Extract, Transform, Load (ETL) overhead to encode points and discretize (polyfill) polygons prior to execution. Furthermore, practical performance is heavily influenced by a tooling gap between highly optimized, industry-backed implementations (e.g., H3, S2) and scientific, equal-area systems (e.g., ISEA4H) that may lack native database integration \cite{juhasz_how_2026, huang_advancing_2024}, despite area-equivalence being a prerequisite for geometrically rigorous, unbiased spatial modeling \cite{juhasz_discrete_2026}.

In this demonstration, we present an interactive dashboard that visualizes the relational throughput and computational tradeoffs of replacing spatial predicates with discrete algebraic hash-joins. Utilizing DuckDB \cite{raasveldt_duckdb_2019}, a vectorized database engine, the dashboard executes real-time spatial joins across multiple DGGS implementations and standard vector baselines. Through interactive global and urban scenarios, the platform allows users to empirically evaluate the upfront encoding overhead against query execution speed, demonstrating the exact data scales at which pre-indexed DGGS datasets surpass traditional vector architectures in performance.

\section{System Architecture}

\begin{figure}[t]
  \centering
  \includegraphics[width=\columnwidth]{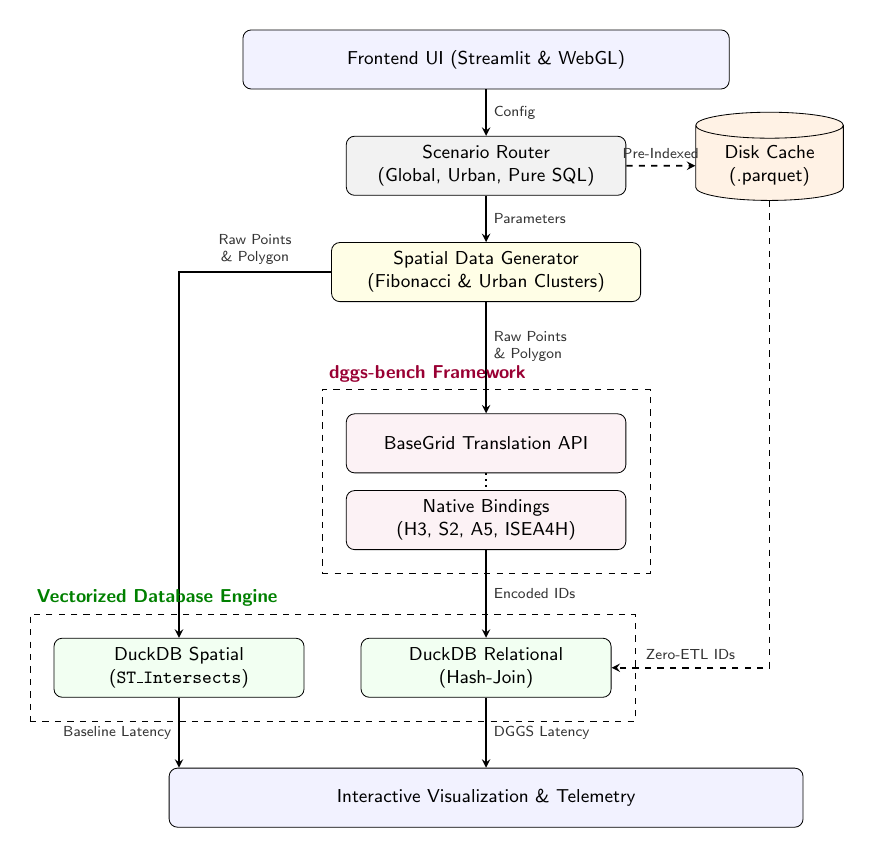}
  \caption{System architecture}
  \label{fig:architecture}
\end{figure}

The goal of the demonstration platform is to empirically compare the latency of traditional point-in-polygon geometric predicates against discrete algebraic hash-joins. As illustrated in Figure~\ref{fig:architecture}, the system is composed of three primary layers: a unified Discrete Global Grid System (DGGS) translation framework, a dual-path vectorized database engine, and an interactive telemetry frontend. Point coordinates are generated via a configurable Spatial Data Generator utilizing global, uniform Fibonacci lattices \cite{juhasz_how_2026, swinbank_fibonacci_2006} and Gaussian-mixture clustering simulating multiple urban centers.

\subsection{Unified DGGS Translation Framework}

A fundamental challenge in evaluating DGGS implementations is the fragmented open-source ecosystem. Production-grade grids such as Uber's H3 \cite{brodsky_h3_2018} and Google's S2 \cite{veach_announcing_2017} are implemented in C and C++, while recent architectures like A5 utilize TypeScript, Python and Rust \cite{palmer_a5_nodate}. Therefore, we utilize the previously developed \texttt{dggs-bench} framework's standardized \texttt{BaseGrid} interface \cite{juhasz_how_2026}, dynamically exposing native grid implementations, including a custom C++ bridge for the scientific ISEA4H grid.

During the ETL phase, \texttt{dggs-bench} performs two crucial operations: polyfilling and encoding. Polygons (e.g., administrative borders or coastlines) are discretized into covering sets of discrete cell identifiers, while point coordinates are encoded into their corresponding cell IDs. This standardizes spatial geometries into flat arrays of integers or strings, preparing them for relational evaluation. Because DGGS are hierarchical, the spatial precision of this discretization is governed by the chosen resolution level, which dictates the physical geographic area of each discrete cell. The \texttt{dggs-bench} framework exposes this resolution as a parameter, allowing the user to explore the trade-off between spatial precision (using smaller, finer cells) and computational throughput (using larger, coarser cells).

\begin{figure}[t]
  \centering
  \includegraphics[width=\columnwidth]{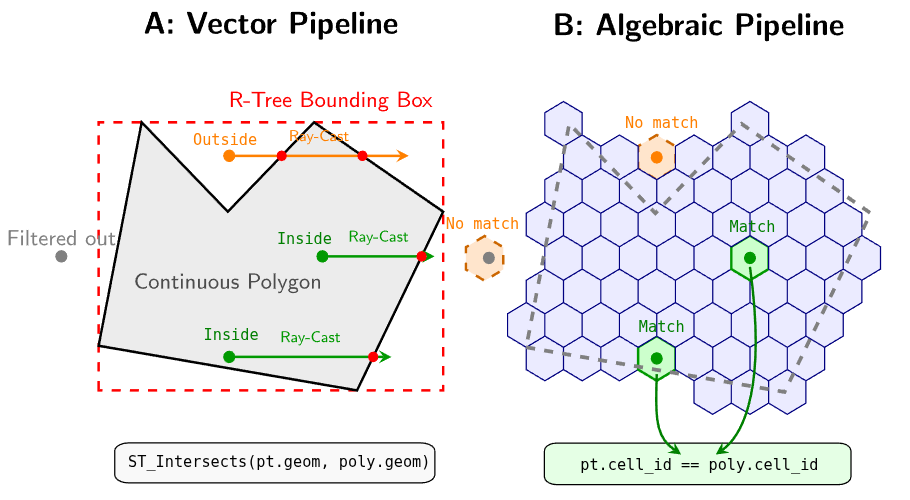}
  \caption{Conceptual comparison of the two approaches. Traditional method using R-trees and ray-casting algorithms (A) and the DGGS approach, reducing containment to an relational hash-join of discrete cell identifiers (B).}
  \label{fig:pipeline}
\end{figure}

\subsection{Relational Execution Engine}

Both pipelines are evaluated within identical environments using DuckDB, a vectorized relational database. As illustrated in Figure \ref{fig:pipeline}A, the traditional spatial baseline relies on the \texttt{ST\_Intersects} predicate. DuckDB utilizes an R-Tree index for a fast bounding-box filter. Any point falling within this rectangle triggers a heavy exact-intersection check via the GEOS geometry engine. This relies on ray-casting algorithms (e.g., computing the even-odd winding number), which scale linearly with the complexity of the polygon boundary. For a polygon with $V$ vertices and $K$ points inside the bounding box, this refinement phase incurs a worst-case time complexity of $\mathcal{O}(K \times V)$.

On the other hand, the DGGS pipeline (Figure~\ref{fig:pipeline}B) operates purely algebraically once input points and polygons are translated into discrete cells. Because the spatial relationships are pre-encoded into the discrete grid identifiers during the ETL phase, the geometric payload is entirely discarded. Point-in-polygon containment is evaluated using a standard relational equi-join (\texttt{pt.cell\_id == poly.cell\_id}). DuckDB evaluates this using an optimized in-memory hash-join. The engine first builds a hash table from the $C$ discrete cells covering the polygon ($\mathcal{O}(C)$), and then probes the $K$ point identifiers against this table in amortized $\mathcal{O}(1)$ time per point. This yields a total execution complexity of $\mathcal{O}(C + K)$. By reducing the spatial problem to relational equality, query performance is fundamentally decoupled from the original polygon's geometric complexity ($V$), bypassing the GEOS engine entirely to unlock the throughput of modern vectorized CPU architectures.

\subsection{Experimental Configuration}

\begin{figure*}[t]
  \centering
  \includegraphics[width=\textwidth]{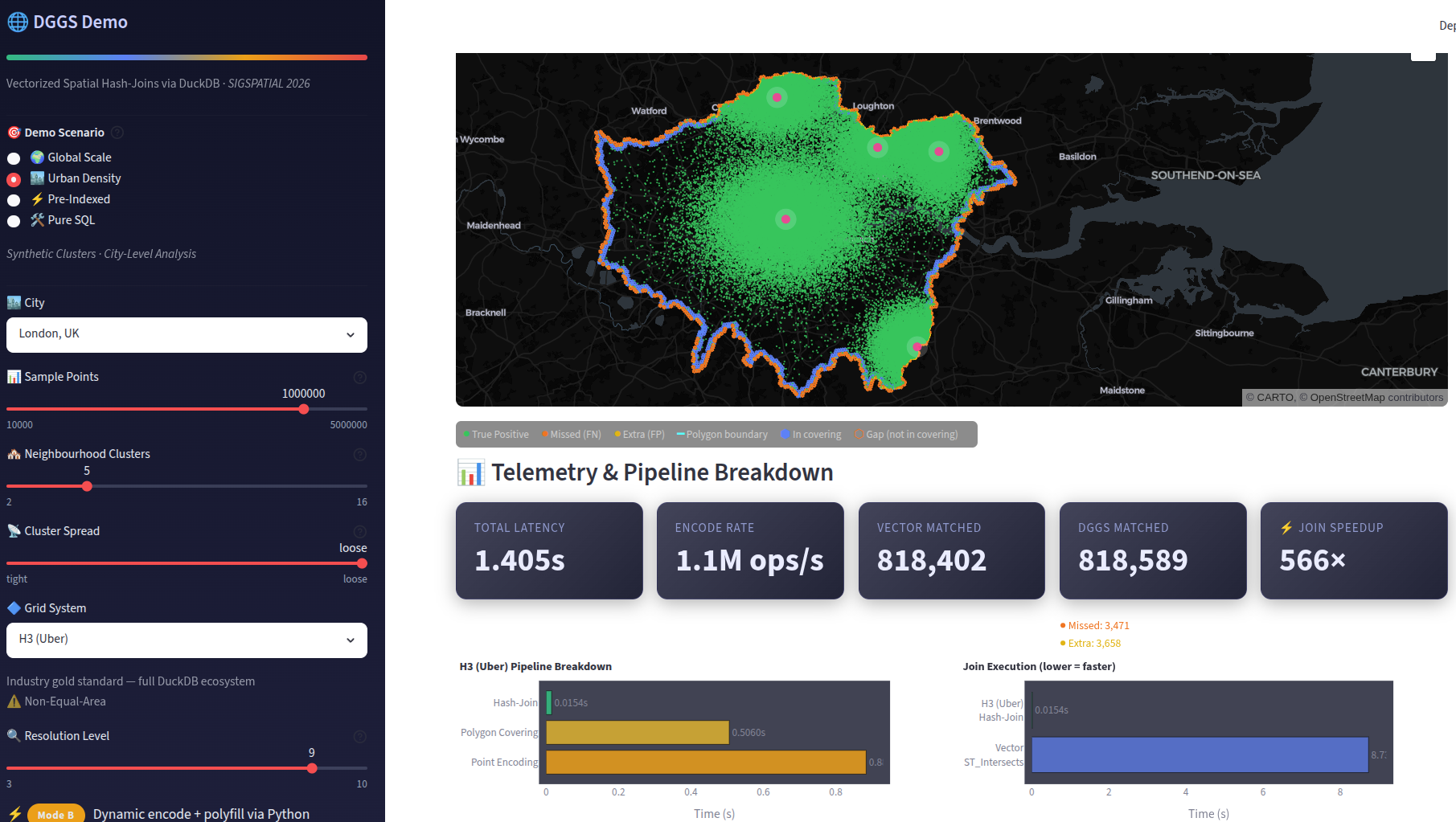}
  \caption{Dashboard in the On-the-Fly scenario (H3, res~10, 1M urban synthetic points).  The telemetry panel decomposes total latency into point encoding, polygon covering, and hash-join phases.}
  \label{fig:dashboard}
\end{figure*}

The dashboard evaluates four DGGS implementations selected to represent a cross-section of current open-source ecosystem maturity (Table~\ref{tab:grids}). Spatial data is generated via two configurable point distributions: a uniform Fibonacci lattice~\cite{swinbank_fibonacci_2006} for global scenarios (up to 5 million  points), and a Gaussian-mixture model that simulates realistic urban clustering for urban scenarios (10k--5M points across configurable city centers). Geographic targets include country-level extents for global benchmarking and  metropolitan areas for urban evaluation. Administrative boundary geometries are retrieved dynamically from OpenStreetMap Nominatim and cached locally.

\begin{table}[h]
\centering
\caption{DGGS implementations and tooling gap.}
\label{tab:grids}
\small
\begin{tabular}{lllll}
\toprule
\textbf{Grid} & \textbf{Type} & \textbf{Equal-Area} & \textbf{DuckDB Ext.} & \textbf{Native Polyfill} \\
\midrule
H3   & Hexagonal     & No & Yes (183 fn.) & Yes \\
A5   & Pentagonal    & Yes & Yes (18 fn.)  & Yes$^*$ \\
S2   & Quadrilateral & No & No           & Python only \\
ISEA4H & Hexagonal  & Yes & No           & C++ bridge \\
\bottomrule
\multicolumn{5}{l}{$^*$\texttt{a5\_geometry\_to\_cells()} available in DuckDB extension v0.9+}
\end{tabular}
\end{table}

\section{Demonstration Scenarios}

The dashboard presents three progressive scenarios that incrementally eliminate computational overhead from the spatial query pipeline. Users interact with a sidebar control console to select the target scenario, DGGS implementation, resolution level, and point count, while the main panel renders a live WebGL map alongside real-time telemetry charts that decompose total latency into its constituent phases. To compare the performance to traditional point-in-polygon,  results for identical conditions (same polygon and point sets) \texttt{ST\_Intersects} results are also presented for each scenario (Figure~\ref{fig:dashboard}).

\subsection{On-the-Fly ETL}

\begin{figure}[t]
  \centering
  \includegraphics[width=\columnwidth]{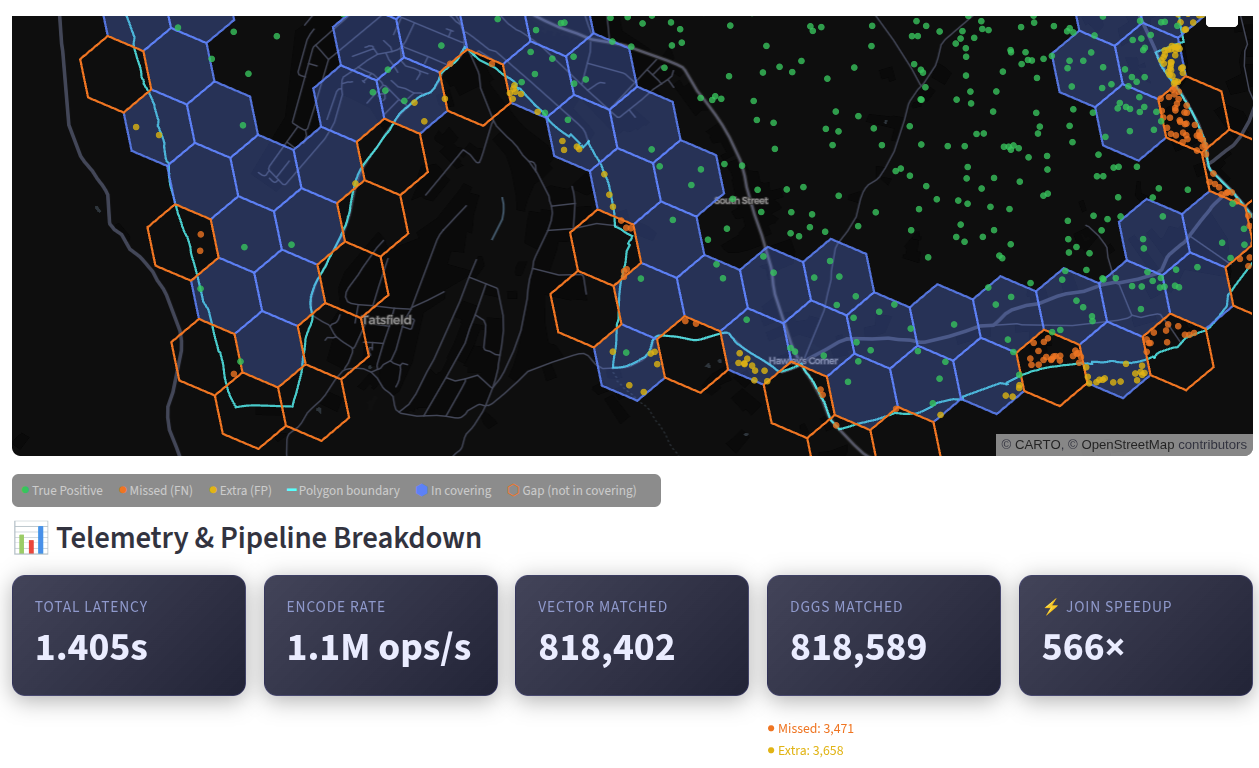}
  \caption{Boundary-edge accuracy for the scenario in Fig.~\ref{fig:dashboard}. Covering cells (blue) and gap cells (orange outline) along London's boundary capture both true and false positives.}
  \label{fig:edge}
\end{figure}

The first two scenarios (Global Scale and Urban Density) execute the full DGGS pipeline at runtime. Point coordinates are encoded into discrete cell identifiers and the target polygon is discretized into a covering set of cells before the hash-join is evaluated. The scenario in Figure \ref{fig:dashboard} shows a $566\times$ speed gain over \texttt{ST\_Intersects} (8.7s vs. 0.015s). The telemetry panel breaks down the total latency by visualizing the pipeline phases as horizontal bars, making the ETL overhead immediately visible. Users observe the tooling gap first-hand: H3 encodes 1~million points in under one second via its optimized C core, while ISEA4H (using a Python-bridged C++ discretizer) requires over 20~seconds for the same workload. The resolution slider allows users to explore the non-linear trade-off between spatial precision and polyfill cost: each additional resolution level multiplies the number of covering cells by the grid's aperture factor (4–7×), so that even a modest increase of two or three levels can grow the covering set by one to two orders of magnitude. Because discrete DGGS tiles only approximate a continuous boundary (see Fig.\ref{fig:pipeline}B), the covering inevitably includes cells extending beyond the boundary (\emph{extra} points) while omitting cells that only partially cover it (\emph{missed} points). As a result, even when the aggregate DGGS count closely matches the \texttt{ST\_Intersects} baseline, the two point sets are not identical: they diverge along the polygon edge. Fig.\ref{fig:edge} illustrates this for the same scenario as Fig.~\ref{fig:dashboard}, color-coding matched, missed, and extra points and reporting the set-level discrepancy.

\subsection{Pre-Indexed Caching}
 
\begin{figure}[t]
  \centering
  \includegraphics[width=\columnwidth]{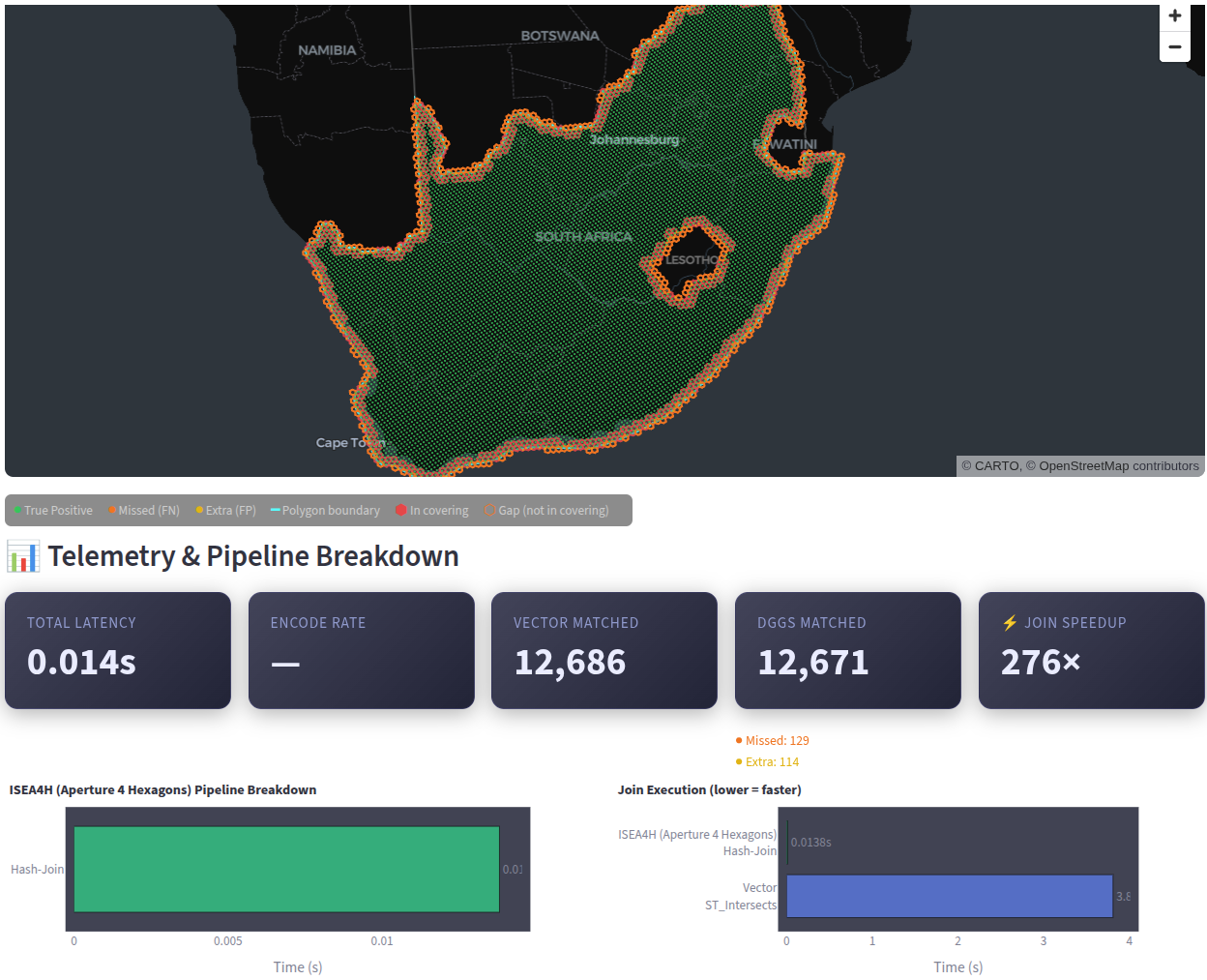}
  \caption{Pre-Indexed Level~2 telemetry: encoding and
    polyfill phases are eliminated.}
  \label{fig:preindexed}
\end{figure}

The Pre-Indexed scenario demonstrates the architectural payoff of distributing spatial data with DGGS cell identifiers already attached. The dashboard loads pre-computed GeoParquet files from disk, containing Fibonacci lattice points (100k, 1M or 5M) encoded across all four grids at two resolutions (coarse and fine) for Brazil and South Africa. Two optimization levels are offered: Level 1, where only input points are pre-encoded, and Level 2, where both points and the polygon covering are pre-generated. This simulates a best-case scenario where geospatial datasets are already available pre-indexed by DGGS identifiers. As shown in Figure~\ref{fig:preindexed}, Level~2 eliminates upfront ETL and leaves only the sub-second join. This convergence holds across all four grids: once pre-indexed, the equal area, scientific but tooling-poor ISEA4H achieves the same join latency as the industry-optimized grids. This proves that current limitations of certain DGGS is an ETL problem, not a query performance problem.

\subsection{Pure SQL Execution}

The final scenario eliminates the Python layer entirely by executing the full spatial pipeline as a single SQL Common Table Expression (CTE) inside DuckDB. This is currently supported for H3 and A5, using their respective community extensions. Users observe that the entire encode--polyfill--join pipeline executes as a single database query with zero serialization overhead, demonstrating that DGGS-native spatial processing is already achievable within standard relational database environments for grids with mature tooling.

\section{Conclusion}
This demonstration provides an interactive platform for empirically evaluating the computational tradeoffs of replacing geometric spatial predicates with algebraic DGGS hash-joins. Through progressive scenarios, users observe that while the current tooling gap imposes significant ETL overhead on equal-area grids, this disparity vanishes entirely when data is pre-indexed: all grid systems converge to sub-second join latency regardless of their underlying mathematical complexity. The \texttt{dggs-bench} framework is available as open-source software at \url{https://github.com/gatorlab-geo/dggs-bench} and the dashboard is also available from \url{https://gatorlab.io}. 

\begin{acks}
During the preparation of this work, the author utilized generative AI tools for two primary purposes. First, AI assistance was used to help edit and refine the manuscript text. Second, the  dashboard was co-developed using autonomous AI agents operating within the AgentLoom dual-helix stabilized framework \cite{boyuan_dual-helix_2026}. The author reviewed, tested, and takes full responsibility for all generated code, experimental results, and the final content of the publication.
\end{acks}

\bibliographystyle{ACM-Reference-Format}
\bibliography{references}


\end{document}